# The Complete Anatomy of the Madden-Julian Oscillation Revealed by Artificial Intelligence

Xiao Zhou[1*], Yuze Sun[1], Jie Wu[2*], Xiaomeng Huang[1*]

1 Department of Earth System Science, Ministry of Education Key Laboratory for Earth System Modelling, Institute for Global Change Studies, Tsinghua University, Beijing, China

2 State Key Laboratory of Climate System Prediction and Risk Management/China Meteorological Administration Key Laboratory for Climate Prediction Studies, National Climate Centre, China Meteorological Administration, Beijing, 100081

## Abstract

Accurately defining the life cycle of the Madden-Julian Oscillation (MJO), the dominant mode of intraseasonal climate variability, remains a foundational challenge due to its propagating nature. The established linear-projection method (RMM index) often conflates mathematical artifacts with physical states, while direct clustering in raw data space is confounded by a "propagation penalty." Here, we introduce an "AI-for-theory" paradigm to objectively discover the MJO's intrinsic structure. We develop a deep learning model, PhysAnchor-MJO-AE, to learn a latent representation where vector distance corresponds to physical-feature similarity, enabling objective clustering of MJO dynamical states. Clustering these "MJO fingerprints" reveals the first complete, six-phase anatomical map of its life cycle. This taxonomy refines and critically completes the classical view by objectively isolating two long-hypothesized transitional phases: organizational growth over the Indian Ocean and the northward shift over the Philippine Sea. Derived from this anatomy, we construct a new physics-coherent monitoring framework that decouples location and intensity diagnostics. This framework reduces the rates of spurious propagation and convective misplacement by over an order of magnitude compared to the classical index. Our work transforms AI from a forecasting tool into a discovery microscope, establishing a reproducible template for extracting fundamental dynamical constructs from complex systems.

# Introduction

The pursuit of science is often a pursuit of definition. From the periodic table in chemistry to the classification of elementary particles in physics, precisely identifying fundamental building blocks has repeatedly catalyzed scientific revolutions by establishing a common framework for understanding and predicting. In the geosciences, a paramount challenge is to objectively define the life cycles of planetary-scale fluid dynamical systems, whose intrinsic complexity, nonlinearity, and propagating nature have long resisted unambiguous decomposition.

The Madden-Julian Oscillation (MJO) exemplifies this challenge. First identified through its distinctive signals in tropical winds and pressure (Madden & Julian, 1971; Madden & Julian, 1972), the MJO is now recognized as the dominant mode of intraseasonal climate variability. This planetary-scale system, characterized by a coupled envelope of deep convection and large-scale circulation anomalies, propagates slowly eastward and modulates a vast spectrum of global weather and climate phenomena (Jiang et al., 2020). It orchestrates monsoon rhythms, energizes tropical cyclone activity, and triggers extreme precipitation events across the globe (Zhang, 2013; Stan et al., 2022; Cheng et al., 2025). Its predictability on subseasonal-to-seasonal timescales makes its accurate monitoring essential for advancing extended-range forecasting (Vitart et al., 2017).

However, the MJO's nature as a propagating mode renders its fundamental definition—and thus its routine monitoring—a uniquely difficult problem. Unlike quasi-stationary phenomena such as the El Niño–Southern Oscillation, trackable via sea surface temperature anomalies in fixed regions (Trenberth, 1997), or compact vortices like tropical cyclones, identifiable by organized eyewall structures (Velden et al., 2006), the MJO is a traveling signal embedded within high-amplitude synoptic noise. Its detection, therefore, cannot rely on local measurements but must instead assess the similarity between the planetary-scale atmospheric state and a set of predefined spatial patterns that capture its essence.

The landmark work of Wheeler and Hendon (2004, WH04) established the operational standard by deriving the Real-time Multivariate MJO (RMM) index from two leading Empirical Orthogonal Function (EOF) modes (Fig. S1 a-b) of filtered, meridionally averaged fields (see Methods). This linear projection framework provided a crucial, globally consistent benchmark. Yet, it suffers from a foundational limitation: it confounds the MJO's mathematical projection with its physical entity. By coupling the system's location and intensity into a single two-dimensional vector derived from fixed, zonally averaged patterns, the RMM index is intrinsically vulnerable to contamination by any non-MJO variability that projects onto these static patterns. This often results in diagnosed "MJO centers" that reside within convectively suppressed regions or trajectories that exhibit unphysical

retrogression—artifacts that undermine physical interpretation and event demarcation. Crucially, this limitation is inherent to the linear, pattern-based approach. Attempts to simply expand the basis by incorporating higher-order EOF modes fail, as these modes lack the canonical planetary-scale, wavenumber-1 structure of the MJO and instead capture noise or other dynamical processes (Fig. S1 c-j). Thus, the WH04 framework is not merely low-dimensional; it is structurally confined to an approximate, and often physically inconsistent, representation.

Given this impasse, a natural proposal would be to extract the MJO's recurrent phases directly from historical data—for instance, by clustering—to refine its life cycle and achieve a more precise description. This intuitive path, however, encounters a more fundamental geometric obstacle: the "propagation penalty." For a propagating phenomenon like the MJO, traditional similarity measures (e.g., Euclidean distance) are vastly more sensitive to spatial displacement than to changes in physical structure. Consequently, in raw data space the same MJO convective-circulation dipole at different longitudes would be treated as distinct classes, preventing any clustering algorithm from identifying the coherent, propagating entity itself (Fig. S2). Defining the MJO objectively therefore requires a new representational paradigm: one that preserves its essential physical attributes (spatial structure, intensity, relative position) while mapping physically similar MJO systems to mathematically nearby vectors, thereby ensuring that similarity reflects physical kinship rather than incidental positional overlap. In recent years, artificial intelligence (AI), especially deep learning, has demonstrated transformative potential in applied fields such as weather forecasting (Pathak et al., 2022; Bi et al., 2023). This raises a deeper question: Can AI move beyond its role as a "prediction tool" to become a "discovery tool," helping to solve foundational theoretical puzzles such as defining the MJO's life cycle?

Guided by this prospect, our study addresses two interrelated scientific questions. First, what is the complete anatomical structure of the MJO's life cycle? Can the long-hypothesized but elusive transitional stages—its organizational growth over the Indian Ocean and its northward shift over the Philippine Sea—be objectively isolated as discrete physical entities? Second, can we establish and test a new research paradigm that turns AI into an engine for theoretical discovery in atmospheric science? Here we attempt what we term the "AI-for-theory" paradigm. Its core aim is to extract physical principles and solve theoretical problems, not to optimize forecasts. It operates through a closed scientific loop: AI-driven feature purification of high-dimensional, nonlinear meteorological fields, followed by physical interpretability checks and traditional diagnostic validation, leading to confirmation, refinement, or generation of testable hypotheses. This work represents a systematic implementation of that paradigm in climate-mode theory. We report that this framework autonomously discovers a coherent, six-phase life cycle for the MJO and, from it, constructs a physically consistent monitoring standard. Our work marks a shift in the role of AI from a "forecasting tool" to an "instrument of discovery," offering a reproducible template for using data-driven methods to uncover the intrinsic organization of complex dynamical systems. Here, we report two key advancements: (1) the first objective, complete anatomical map of the MJO's six-phase life cycle, discovered by an AI trained to overcome the propagation penalty; and (2) a new, physics-coherent monitoring framework that, derived from this anatomy, eliminates long-standing artifacts in MJO tracking.

# Results

## 1. An AI-Discovered Anatomy of the MJO Life Cycle

To address the most fundamental question: what are the recurrent physical states that constitute the MJO's life cycle? By applying a nonlinear self-organizing map (SOM, see Methods) clustering algorithm to the learned "MJO fingerprints" — the latent vectors that encode physical feature similarity — we objectively identify six recurrent patterns. These six canonical phases form a complete anatomical map of the MJO that is directly rooted in, and extends, the classical understanding (Fig. 1; see Fig. S3 for detailed structures). The specific clustering procedure is described in detail in the following section.

Ordered by the longitude of their convective center, the six phases self-assemble into a physically coherent progression. Each exhibits the canonical baroclinic structure of the MJO—coupled deep convection and a first-baroclinic-mode zonal wind response. Critically, our objective taxonomy does not emerge in a vacuum; it refines and completes the classical EOF view:

- **Phase 3 (Maritime Continent) and Phase 5 (Western Pacific)** align closely in spatial structure with the first and second classical EOF modes, respectively, capturing the mature MJO signal in these long-established regions.

- **Phase 1 (Eastern Africa) and Phase 6 (Central Pacific)** unambiguously describe the large-scale suppressed convective states that are approximately anti-phasic to the active centers. They are not perfect mirror images, offering a more nuanced description of the MJO's life-cycle symmetry.

More significantly, our framework objectively isolates two long-hypothesized but previously undefined transitional stages as discrete, co-equal phases:

- **Phase 2 (Indian Ocean)** provides the first spatially coherent, objective signature of MJO growth over the Indian Ocean. It manifests as the large-scale organization and rapid intensification of deep convection, setting the stage for subsequent maturity. In the linear EOF framework, this rapidly evolving, structurally nascent signal is obscured by the dominant mature patterns that are optimized to explain maximum variance.

- **Phase 4 (Philippine Sea)** unambiguously captures the pronounced northward shift of the convective center over the Philippine Sea—a key two-dimensional structure arising from interaction with the monsoonal background flow, a feature entirely erased by the meridional averaging inherent in the classical index.

Together, these six phases constitute a highly representative "state library" of the MJO. Approximately 75% of all days in the historical record are classified into one of these six canonical phases, indicating that they encapsulate the vast majority of physical states during active MJO periods. This stands in indirect contrast to the ~23% combined variance explained by the first two classical EOF modes (Fig. S1 a-b), suggesting that our method, through its similarity-preserving latent representation, more effectively aggregates the dynamical information coherent with the MJO.

The six-phase anatomy is a robust property of the MJO's intrinsic dynamics, not an artifact of the specific clustering algorithm. The same set of phases emerges from linear K-Means clustering applied to the same latent vectors (Fig. S4). The Self-Organizing Map itself was configured using an objective error-minimization criterion (Fig. S5), and it identified two additional, non-canonical clusters (Fig. S6). These two clusters violate the necessary condition for MJO identification: a planetary-scale, coherent, wavenumber-1 convective–circulation couplet. Their exclusion therefore follows directly from the physical definition of the MJO, rather than subjective preference.

In summary, the "AI-for-theory" paradigm distills from data the first complete anatomical map of the MJO—one that inherits, refines, and completes the classical view. The objective identification of Phase 2 (growth) and Phase 4 (northward shift) as canonical phases resolves long-standing definitional ambiguities and provides precise observational targets for diagnosing the MJO's predictability and impacts.

## 2. Learning an "MJO Fingerprint" to Overcome the Propagation Penalty

The six canonical phases presented in Section 1 were derived by clustering daily atmospheric states. However, clustering in raw data space is fundamentally challenged by the MJO's propagating nature. Traditional similarity metrics (e.g., Euclidean distance) are overwhelmingly sensitive to zonal displacement—the "propagation penalty" (Fig. S2)—causing a single MJO event at adjacent longitudes to appear distinct and rendering direct clustering ineffective amidst high-frequency noise.

To overcome this, we designed the **PhysAnchor-MJO-AE** model to learn a latent representation guided by three principles: information retention, similarity preservation, and uniqueness. We adopt an autoencoder (AE) architecture to ensure the latent vector retains essential physical attributes by accurately reconstructing the input. To achieve similarity preservation, we draw inspiration from word embeddings in natural language processing: just as semantically similar words are mapped to nearby vectors, we train the encoder to map physically similar MJO states to geometrically close latent vectors (see Methods). This is realized through a self-supervised task where the model reconstructs a day's state from a short sequence of prior days, thereby learning the MJO's contextual "grammar" and organizing the latent space to reflect physical kinship. Uniqueness is inherently enforced by the deterministic encoder and the consistency of the learned dynamics. Thus, the learned representation ensures that distance in the latent space reflects similarity in MJO physical features, rather than proximity in geographic location, laying a principled foundation for clustering based on dynamical resemblance.

The implementation of **PhysAnchor-MJO-AE** model (see Methods) is trained to reconstruct the third day from three consecutive filtered fields. This window optimally balances context while confining MJO propagation to a local sector (~10° longitude). The encoder learns to distill a compact 1024-dimensional latent vector (Fig. 2b) that captures the coherent signal, mapping adjacent observations of the same event to highly similar vectors—a true similarity-preserving embedding. Validation confirms this representation is accurate (Fig. 2c), preserves classical monitoring information (Fig. S7), and supports meaningful linear operations for clustering (Fig. S8). Thus, PhysAnchor-MJO-AE transforms noisy snapshots into clean "MJO fingerprints," providing the principled latent vectors that enabled the objective phase discovery in Section 1 and circumventing the historical limitations imposed by the propagation penalty.

## 3. A Physics-Coherent Framework for Defining and Tracking MJO Events

Accurately diagnosing the daily state of the Madden-Julian Oscillation requires unambiguous separation of its signal from ubiquitous synoptic noise. The classical RMM index (WH04) approaches this by projecting observed fields onto two fixed, zonally averaged EOF patterns. Mathematically, this projection calculates regression coefficients (principal component time series) that are proportional to the covariance between the daily data and the patterns. Within the linear decomposition framework, this methodology constituted an optimal description of the MJO at the time of its development and, most importantly, established the first unified, objective standard for global MJO monitoring, profoundly advancing the field. However, this covariance-based approach is inherently sensitive to absolute amplitude: a strong, non-MJO convective event (e.g., a tropical cyclone or monsoon depression) will generate a large projection coefficient, even if its spatial structure differs significantly from the canonical MJO. Consequently, the index confounds true MJO variability with high-amplitude, incoherent noise. This fundamental sensitivity contaminates both elements of the index—the derived "location" (phase angle) and the "amplitude"—leading to trajectories that can jump westward or place the MJO center in non-convective regions, as quantified below.

Our framework overcomes this core limitation through a decoupled and structurally grounded methodology. First, we diagnose location and intensity independently, basing both on the full 2D structure of the six canonical phases. Instead of a covariance-based projection, we compute the spatial pattern correlation between the observed fields and each phase. This correlation metric is normalized by the local variance of both the observation and the phase pattern, making it inherently insensitive to absolute amplitude differences. It quantifies purely the geometric similarity in spatial structure, effectively filtering out the influence of strong but structurally dissimilar non-MJO features. The MJO's daily longitude is then calculated as a weighted average of the phases' pre-defined characteristic system centers—defined as the mean of the convective (OLR) and wind-field anomaly centers—where the weights are the normalized pattern correlation strengths. Its intensity is independently computed from a weighted root-mean-square of projections, preserving the system's energy but now using weights derived from the robust correlation measure.

To manage real-world complexity—such as the coexistence of a decaying system and a nascent one—we introduce a physically-based disambiguation rule. This rule is grounded in the observed behavior of MJO events: while multiple convective envelopes can exist simultaneously, operational monitoring and physical understanding typically focus on the most dominant, coherent signal at a given time. Our algorithm identifies the single best-matching phase and restricts the location calculation to phases within the same longitudinal hemisphere. This ensures that only one dominant convective envelope is tracked continuously. The rule allows for event transition: tracking switches to a new, growing system only when its pattern correlation with the canonical phases surpasses that of the previously tracked, decaying system. This design provides clear segmentation of distinct events and aligns with the objective recognition of competing life cycles in nature.

Guided by this robust diagnostic approach, we define an MJO event through three physical principles: structural coherence, sustained strength, and realistic propagation. An event is identified when, for at least 15 consecutive days, the daily state: (1) exhibits a maximum pattern correlation with any canonical phase above an objective threshold ($r > 0.36$, defined by the 25th percentile of the historical distribution; Fig. S9); (2) has an intensity index $> 1.0$;

and (3) shows net eastward propagation. This definition moves beyond the noise-sensitive amplitude thresholds of the classical approach (see Methods).

**Long-term performance and physical coherence**

We first evaluate the framework's performance across decades. Figure 3 juxtaposes the diagnosed trajectories of the MJO's convective center from both the new method and the classical RMM index, overlaid on the observed propagation of OLR anomalies. For a spatially continuous and fair comparison, the discrete phases of the WH04 index were mapped to a representative central longitude based on its canonical patterns. Quantitatively, over the entire 40-year period, the classical method produces physically implausible propagation (spurious westward jumps or eastward propagation faster than 10° longitude/day) on 23% of all MJO event days, and places the diagnosed convective center within convectively suppressed (positive OLR anomaly) regions 17% of the time. In stark contrast, our new method reduces these error rates to merely 0.24% and 3%, respectively. Visually, the new framework produces trajectories that are physically consistent across epochs: they closely track the core of eastward-propagating convective anomalies (symbols within negative OLR), maintain realistic propagation speeds (primarily <10° longitude/day), and yield cleanly segmented event life cycles. In stark contrast, the classical index frequently places the MJO center within convectively suppressed regions (symbols in positive OLR), exhibits sporadic unphysical westward jumps, and merges distinct convective envelopes. This multi-decadal comparison demonstrates that the decoupled, pattern-similarity-based approach fundamentally overcomes the noise susceptibility inherent in linear projection.

**Case studies revealing systemic flaws and resolution**

To dissect the specific mechanisms of failure and improvement, we examine three archetypal events (Fig. 4). The first case (Fig. 4a-c) highlights the problem of spurious propagation and location error. While the new index tracks steady eastward movement co-located with the convective core, the RMM index shows a clear retrogression (westward jump) absent in observations and, on a key date, locates the MJO in a non-convective region. The second case (Fig. 4d-f) underscores the failure in representing intensity evolution. The new intensity index accurately reflects the relative strength of two convective peaks (ratio ~1.7:1), mirroring the OLR anomaly contrast, whereas the RMM amplitude remains nearly flat. The third case (Fig. 4g-i) exposes the inability to distinguish concurrent events. When two independent MJO systems are separated by a suppressed corridor, the new framework cleanly segments them using its physically-based disambiguation rule, while the classical index erroneously merges them, placing the tracked center in the suppressed zone between them.

Collectively, these results demonstrate that the classical index's coupled, low-dimensional representation is inherently prone to generating physically implausible states. Our framework, derived from and validated against the objective MJO anatomy, systematically resolves these flaws by decoupling location and intensity, prioritizing structural coherence, and applying physically sensible rules, delivering a robust and reliable standard for operational monitoring and research.

## 4. A Standardized Template for MJO Lifecycle Analysis

The objective phase anatomy and the decoupled monitoring framework together provide the foundation for a comprehensive, physics-based depiction of MJO activity. We synthesize these components into a standardized analysis template that transcends the limited, abstract perspective of the classical phase diagram, offering a unified and intuitive summary of individual MJO events (Fig. 5). This template is designed not merely as a visualization tool, but as an integrated diagnostic environment that bridges derived indices with raw observational context, ensuring physical interpretability at every step.

This template presents two complementary, co-equal perspectives in a single view:

1. **Phase and Intensity Evolution (Fig. 5a):** This panel chronicles the event's core evolution by plotting its daily state in a physical longitude-intensity space. The trajectory traces a coherent path where the symbol and color at each point denote the dominant canonical phase, thereby simultaneously visualizing the system's location, strength, and structural identity. This direct representation in physical coordinates contrasts with the abstract, mathematically coupled coordinates of the classical RMM phase space (inset), where identical points can correspond to vastly different physical states.

2. **Dynamical and Convective Context (Fig. 5b):** This panel grounds the diagnostic view in the observed atmosphere. A Hovmöller diagram displays the propagation of large-scale OLR and zonal wind anomalies, with the MJO's daily location and phase—diagnosed by our framework—overlaid using the same symbology as panel (a). The **precise alignment** of this trajectory with the heart of the observed convective disturbance provides direct, visual validation that our index tracks the **physical entity itself**, not a mathematical abstraction susceptible to noise. It contextualizes the event's evolution within the broader tropical wave spectrum.

The power of this integrated template lies in its simultaneity and consistency. By combining the state evolution plot with its environmental context, it delivers a more complete and physically transparent summary than the classical phase diagram. It seamlessly communicates the real-time state, historical trajectory, canonical phase, and underlying dynamics in a single, coherent visualization. This design directly addresses the core limitation of classical methods in accurately pinpointing MJO location, providing a more robust tool for MJO identification and monitoring.

We propose this physics-based template as a new standard for operational monitoring, diagnostic research, and model evaluation of the MJO. Its adoption can facilitate clearer communication, more consistent analysis, and rigorous comparison across the community, effectively bridging the gap between fundamental dynamical insight and operational utility. The template encapsulates the core advance of our work: moving from a low-dimensional, abstract index to a high-dimensional, physically anchored representation of the climate system's dominant mode of subseasonal variability.

# Discussion

## 1. An Objective Anatomy: Completing the MJO Life Cycle

Our most fundamental finding is an objective, six-phase anatomical map of the Madden-Julian Oscillation, discovered autonomously by our AI-for-theory framework. This structure provides a long-missing, unified taxonomy for the phenomenon, systematically refining and—critically—completing the classical view derived from linear EOF analysis. While Phases 3 (Maritime Continent) and 5 (Western Pacific) correspond to the mature convective centers of the traditional index, and Phases 1 and 6 describe their suppressed counterparts, our analysis provides a more nuanced physical description. Notably, the suppressed phases are not perfect mirror images of the active ones, challenging the implicit antisymmetry often assumed in simplified linear models.

The pivotal advance is the formal isolation of two transitional stages as discrete, canonical phases, whose physical rationality is underpinned by robust dynamical processes and consistent with existing meteorological understanding. Phase 2 (Indian Ocean) robustly captures the MJO's growth—the large-scale organization and intensification of deep convection following moisture preconditioning (Li et al., 2015). Its convective enhancement is driven by the synergy of equatorial wave coupling and air-sea feedback: the resonance of equatorial Kelvin waves and Rossby waves excites low-level convergence here (Kiladis et al., 2009), superimposed with a "convection-moisture" positive feedback loop for energy accumulation (Adames & Maloney, 2021). Meanwhile, sea surface temperature anomalies in the Indian Ocean Warm Pool serves as key prerequisites for its smooth development (Zhao et al., 2013). The classical EOF framework, optimized to extract maximum variance, inherently blurs this structurally evolving, often lower-amplitude signal in favor of the dominant mature patterns. Phase 4 objectively defines the MJO's northward shift over the Philippine Sea, a key structural signature of its interaction with the monsoonal flow (Hsu et al., 2011). This phase is a critical dynamical link to high-impact weather: it provides a favorable environment for western Pacific tropical cyclogenesis (Li & Zhou, 2013) and its off-equatorial heating pattern efficiently forces extratropical teleconnections (Adames & Wallace, 2014; Stan et al., 2017). While the classical RMM index erases this two-dimensional structure, our framework preserves this physical nexus, enabling precise diagnosis of how the MJO connects its core dynamics to remote weather extremes.

The implications are immediate and substantive, highlighting their significant scientific value. Critically, these are independent phases: neither Phase 2 nor Phase 4 can be derived as a linear combination of the two canonical WH04 EOF modes. Phase 2 provides a precise observational target for diagnosing the MJO's often poorly simulated initiation and growth in dynamical models, directly addressing a key predictability bottleneck—this fills the long-standing gap in MJO initiation mechanism research where the transitional stage could not be accurately localized. Phase 4 directly encapsulates the dynamical link to high-impact weather: its off-equatorial heating pattern is a known catalyst for Western Pacific tropical cyclogenesis and a potent forcing for mid-latitude Rossby wave trains (Zhao et al., 2015). By formalizing the MJO-monsoon coupling process as a discrete phase, Phase 4 offers a quantifiable observational target for precisely revealing the MJO precursor mechanisms of high-impact weather, which was previously confounded with Phase 5 signals by traditional methods. Thus, our anatomical map does not merely add detail; it bridges core dynamics with consequential impacts, offering a foundational template for mechanistic and predictive studies.

## 2. A New Lens for Discovery: The "AI-for-Theory" Paradigm

The pathway to this discovery exemplifies a new research paradigm—one we term "AI-for-theory." Moving beyond AI's predominant role as a black-box predictor, this approach employs AI as a principled microscope for scientific discovery, with the core objective of extracting physical principles and solving foundational theoretical problems, rather than optimizing forecasts. The paradigm operates through a structured, closed-loop methodology: AI-driven feature purification distills informative representations from high-dimensional data; physical interpretability and diagnostic validation ground these patterns in established dynamical knowledge; and the resulting insights are used to confirm, refine, or generate testable hypotheses. This study provides its first complete instantiation: by learning a similarity-preserving embedding and deriving the MJO's six-phase anatomy, we transform AI from a forecasting tool into a discovery tool that resolves a long-standing definitional puzzle in climate dynamics.

This paradigm's power is illustrated by the model's autonomous distillation of the MJO's dynamical blueprint. It determined that encapsulating the MJO's coherent essence required a latent representation of ~1024 dimensions—a data-driven measure of its effective dynamical complexity. The model acts as an abstraction engine, compressing the input by extracting the recurring, defining structural relationships of the MJO (e.g., the baroclinic wind couplet). Crucially, it structures the latent space so that positional and dynamical proximity are aligned. This meaningful geometry—where distance reflects similarity—emerged autonomously from the reconstruction task, showcasing how a tailored learning objective can force AI to extract physically interpretable insight.

The success with the MJO opens a clear path for applying this paradigm to other complex geophysical systems where definitions remain contentious. Could a similar framework objectively identify the distinct "flavors" and complete life cycles of ENSO, moving beyond fixed regional indices? Might it characterize the genesis, maturation, and decay of atmospheric blocking events, or delineate the life cycle of oceanic mesoscale eddies? The core strength of AI-for-theory lies in its ability to learn structure-aware, similarity-preserving representations, offering a rigorous new lens for the fundamental scientific task of regime identification in systems dominated by propagation, translation, or morphing shapes. By establishing a reproducible workflow from data to physical taxonomy, this paradigm provides a template for theory-driven AI across the geosciences.

## 3. From retrieved taxonomy to AI-native representation

Our current framework, while transformative, has intentional limitations that point toward future research. To ensure a direct and fair comparison with the established benchmark, we adopted the foundational variable selection and spatial filtering of the WH04 scheme. Future work could investigate whether incorporating additional fields (e.g., column-integrated moisture, diabatic heating) or using higher-resolution data yields a richer characterization of multi-scale interactions within the MJO envelope. Furthermore, our monitoring algorithm, while robust and physics-coherent, remains a "retrieval" method based on similarity to predefined phases discovered from a historical period.

A more ambitious, end-to-end paradigm lies ahead: training models to directly and optimally isolate the MJO signal from raw or minimally processed data in real time. Such an AI-native representation would simultaneously diagnose location, intensity, and spatial configuration as intrinsic, dynamic outputs of the model, moving beyond supervised retrieval toward a learned, generative model of the phenomenon itself. This could lead to adaptive indices that evolve with the climate system and provide a crucial step toward more intelligible and capable 'digital twins' for subseasonal prediction.

## 4. Concluding remarks

In summary, we have used deep learning not as a black-box predictor, but as a theory-discovery engine to advance the physical understanding of a major climate mode, delivering its most complete anatomical picture. The resulting framework—with its objective phases, decoupled index, and principled event definition—offers a new standard for the community. A key design philosophy is the separation of the computationally intensive, one-time AI-driven discovery phase from its lightweight, transparent application. To enable immediate adoption, we publicly release the six canonical phase patterns and an open-source software package that performs real-time monitoring and generates the standardized analysis templates (Fig. 5). This turnkey approach minimizes the barrier to entry, allowing the community to leverage the new anatomy without requiring specialized AI expertise.

Our work represents an attempt to advance the methodology of the climate sciences. Historically, the field has progressed through successive paradigms: observational characterization, theoretical distillation, and numerical simulation. The approach demonstrated here—using AI to help illuminate intrinsic structure and aid in extracting governing principles from data—points toward a complementary, emerging direction: AI-assisted scientific discovery. In this framework, artificial intelligence aims to transcend its role as a forecasting tool and serve as an instrument of inquiry, seeking to interface with complex systems in order to generate testable physical hypotheses and objective taxonomies. We suggest that the "AI-for-theory" approach could expand the ways in which we interrogate the organized complexity of the natural world, offering a promising lens for future geoscientific research.

# Methods

## Data

This study utilizes the core MJO diagnostic variables established by Wheeler and Hendon (2004): outgoing longwave radiation (OLR) and zonal winds at 850 hPa (U850) and 200 hPa (U200). Daily U850 and U200 data are sourced from the NCEP–DOE AMIP-II Reanalysis (NCEP2; Kanamitsu et al., 2002), and OLR data from the National Oceanic and Atmospheric Administration (NOAA; Liebmann & Smith, 1996). The analysis period spans 1979–2020. All data are interpolated to a 2.5° × 2.5° grid. To isolate the intraseasonal MJO signal, raw daily fields are first filtered using a 20–90-day Lanczos bandpass filter—a foundational step consistent with the classical WH04 methodology.

The WH04 index is derived by meridionally averaging (15°S–15°N) these filtered fields, performing an EOF decomposition, and retaining the first two principal modes as the canonical MJO phases. In our new framework, we build upon this common 20–90-day filtered data. For subsequent calculations (e.g., pattern correlation with canonical phases), we apply an additional zonal wavenumber 1–3 spectral filter to the intraseasonal anomalies. This extra step refines the signal by retaining only the planetary-scale component, thereby reducing contamination from smaller-scale convective noise and enabling a more precise, structure-based assessment of MJO activity.

## Inspiration from Word Embeddings

The core idea of our approach is inspired by word embeddings in natural language processing (NLP). In NLP, words must be converted into numerical vectors for computational processing. The most effective encodings map semantically similar words to nearby points in a high-dimensional latent (embedding) space—for example, the vectors for "cat" and "dog" are closer than those for "cat" and "car." This geometric organization allows models to efficiently capture semantic relationships. A celebrated property of such spaces is their linear regularity: semantic analogies can often be solved via vector arithmetic (e.g., $\vec{vec}[\text{"king"}] - \vec{vec}[\text{"man"}] + \vec{vec}[\text{"woman"}] \approx \vec{vec}[\text{"queen"}]$).

We adopt this paradigm for the MJO. We treat each day's multivariate atmospheric state (U850, U200, OLR) as a "climate word." Our model is designed to learn an encoder that maps these states into a similarity-preserving embedding space, where physically akin and positionally proximate MJO systems are mapped to nearby vectors. In this learned "climate embedding" space, the life cycle of an MJO event traces a continuous trajectory, and recurrent dynamical regimes emerge as distinct clusters. This representation directly circumvents the "propagation penalty" that plagues clustering in raw data space, enabling the objective discovery of the MJO's canonical phases.

## Model Architecture: PhysAnchor-MJO-AE

Our model, PhysAnchor-MJO-AE, is designed to learn a latent representation that satisfies three core principles: information retention, similarity preservation, and uniqueness. It is built on a Transformer-based encoder-decoder framework, trained to understand the MJO from short sequences of atmospheric states.

**Input, Output, and Sequence Length.**

The model input is a sequence of three consecutive daily maps (days t−2, t−1, t) comprising three preprocessed filtered variables: U850, U200, and OLR. Its objective is to reconstruct the detailed spatial fields of the target day (t). The three-day window was selected after testing intervals of 2–7 days, yielding the minimum reconstruction loss. Physically, this window confines the MJO's propagation (~5° longitude/day) within a local longitudinal sector, maximizing the coherence of its dynamical evolution while providing sufficient context for learning.

**Encoder–Decoder Structure.**

A convolutional neural network (CNN) first extracts spatial features from each daily map. A subsequent Transformer encoder (Vaswani et al., 2017) stack (3 layers, 8 attention heads) integrates these features across the temporal sequence via self-attention, identifying and linking the coherent, evolving components of the MJO. The decoder mirrors this architecture, mapping the encoder's output back to the full spatial domain to perform reconstruction.

Learning Objective and Principle Realization.

The model is trained in a self-supervised manner using a mean-squared-error (MSE) loss between the reconstructed and true fields of day t.

- Information retention is achieved by the autoencoder (AE) design: accurate reconstruction forces the latent vector to retain essential physical attributes of the input.

- Similarity preservation is realized through the sequence reconstruction task. Drawing inspiration from word embeddings, this objective teaches the encoder to map physically similar and positionally proximate MJO states to geometrically close latent vectors. By learning the MJO's contextual "grammar," the model organizes the latent space to reflect physical kinship.

- Uniqueness is inherently enforced by the deterministic nature of the encoder and the consistency of the learned dynamical relationships.

**Latent Representation.**

The output of the Transformer encoder is a 1024-dimensional latent vector for each input sequence. This vector serves as a similarity-preserving "MJO fingerprint" and provides the input for the subsequent clustering analysis that identifies the canonical phases.

## Clustering Procedure

To objectively identify the recurrent, dynamical states constituting the MJO life cycle, we perform clustering on the learned latent vectors derived from the contextual AI model, which encode MJO dynamical states such that geometrically close vectors correspond to physically similar and positionally proximate MJO systems.

**Primary Clustering via Self-Organizing Map (SOM).**

The daily 1024-dimensional latent vectors serve as input to an unsupervised SOM algorithm. The SOM projects these high-dimensional vectors onto a discrete, two-dimensional grid of nodes while preserving their topological relationships. This non-linear method is well-suited for discovering the intrinsic, low-dimensional manifold of a complex dynamical system. The optimal number of clusters was determined by minimizing a combined error metric of quantization error and topographic error across multiple SOM configurations. This objective criterion identified a 4×2 (8-node) topology as optimal for our dataset (Fig. S5). The final clustering yielded eight nodes.

**Phase Selection and Validation.**

From the eight SOM nodes, we selected the six that exhibit the canonical, planetary-scale wavenumber-1 structure and the characteristic coupled convection-circulation signature physically defining the MJO. The remaining two nodes, which lacked this coherent, large-scale structure, were excluded as non-canonical states (Fig. S6). The robustness of the resulting six-phase taxonomy was confirmed by applying linear K-Means clustering (with k=6) to the same latent vectors, which produced a highly consistent set of spatial patterns (Fig. S4). The six canonical phases are ordered by the longitude of their characteristic system center to form the complete life cycle presented in Fig. 1.

## New Index Calculation

The monitoring framework calculates two independent metrics—longitude and intensity—from the observed atmospheric fields, based on their similarity to the six canonical MJO phases.

**Pattern Correlation and Phase Assignment.**

For each day, we compute the spatial pattern correlation coefficient $r_i$ between the observed, filtered fields (U850, U200, OLR) and each canonical phase $i$ ($i=1,…,6$) within the analysis domain (15°S–15°N, 30°E–210°E). The phase with the maximum correlation $r_{max}$ is assigned as the dominant phase for that day. This correlation-based measure is inherently normalized, focusing on structural similarity and minimizing sensitivity to absolute amplitude variations from non-MJO disturbances.

**Weight Calculation via SoftMax Transformation.**

To derive a smooth and positive weight for each phase, we transform the correlation coefficients $r_i$ using a SoftMax function. This design ensures that even negative correlations (indicating an anti-correlated structure) contribute minimally but continuously to the weighted average, avoiding abrupt discontinuities that would arise from setting them to zero. The transformation is defined as:

$$w_i = \frac{\exp(f \cdot r_i)}{\sum_{i=1}^{6} \exp(f \cdot r_i)}$$

where $f=3$ is a smoothing factor. This factor, calibrated against historical data, optimally balances the sensitivity to high correlations (ensuring accurate phase assignment) with the need for smooth transition in weights when correlations are weak or negative.

**Longitude Calculation.**

The daily MJO longitude ($\lambda_{MJO}$) is derived as a weighted average of the characteristic system centers ($\lambda_i$) of all six phases. The characteristic center for each phase is defined as the mean of its convective (OLR) anomaly center and its baroclinic wind (U850/U200) anomaly center.

$$\lambda_{MJO} = \sum_{i=1}^{6} w_i \cdot \lambda_i$$

**Intensity Calculation.**

The daily MJO intensity index ($I_{MJO}$) is computed independently as a weighted root-mean-square measure of projection amplitudes, scaled to be directly comparable in magnitude to the classical RMM amplitude:

$$I_{MJO} = \alpha \cdot \sum_{i=1}^{6} w_i \cdot \alpha_i^2$$

Here, $\alpha_i$ is the projection amplitude (regression coefficient) of the observed fields onto phase $i$. The scaling factor $\alpha=3$ is introduced because our canonical phases are normalized. This factor does not alter the relative evolution of intensity but conveniently places $I_{MJO}$ on a similar numerical scale to the WH04 index, facilitating direct comparison for users.

**Physically-Based Disambiguation Rule.**

To ensure the unambiguous tracking of a single, dominant MJO event at any given time—particularly when convective signals exist in both the Indian Ocean and Pacific sectors—we implement a disambiguation rule based on longitudinal coherence.

The rule is applied as follows: after computing the pattern correlation $r_i$ for all six phases, we identify the phase with the maximum correlation. This dominant phase defines the active longitudinal hemisphere for the day:

- If the dominant phase is Phase 1 or 2 (primarily residing in the Eastern Hemisphere, e.g., 30°E–120°E), the weights $w_i$ for phases in the opposing hemisphere (Phases 5 and 6, primarily in the Pacific sector) are explicitly set to zero before the longitude and intensity calculations.

- Conversely, if the dominant phase is Phase 5 or 6, the weights for Phase 1 and 2 are set to zero.

This mechanism ensures that the calculated longitude and intensity metrics are derived solely from the most structurally similar, and likely physically connected, set of phases within one longitudinal domain. It prevents the index from averaging the locations of two potentially independent convective envelopes, which would yield a physically meaningless central longitude situated in a convectively suppressed region between them.

Furthermore, this rule governs event transitions. An MJO event is considered ongoing as long as a consecutive sequence of days is dominated by phases within the same hemisphere. The event is considered terminated, and tracking may switch to a new initiating event in the opposite hemisphere, only when the correlation strength associated with the original hemispheric domain decays below a sustained threshold, allowing a new, growing signal to become dominant. This provides a clear, physically based segmentation of distinct MJO life cycles.

**Event Definition.**

An MJO event is identified when the following criteria are met consecutively for at least 15 days:

1. **Structural Coherence:** $r_{max} \geq 0.36$. This threshold is the 25th percentile of the historical daily $r_{max}$ distribution (Fig. S9). It is objectively chosen to be consistent with our clustering results, which identified ~75% of days as belonging to a canonical phase (Fig. 1), implying that ~25% of states are unstructured with respect to the MJO template.

2. **Sustained Strength:** The daily intensity index must satisfy $I_{MJO} \geq 1$. The scaling factor α ensures that $I_{MJO}$ is on a directly comparable scale to the classical RMM amplitude, making this threshold physically consistent with established convention.

3. **Realistic Eastward Propagation:** The diagnosed daily longitude must exhibit net eastward motion over the event duration. To quantify this, a daily zonal phase speed $c$ is calculated using a centered difference:

$$c_t = \frac{\lambda_{t+1} - \lambda_{t-1}}{2}$$

where $\lambda$ is the MJO longitude in degrees east. The mean phase speed over the event must be positive, and individual daily speeds are typically less than 10°longitude per day, consistent with the observed propagation of the MJO.

This tripartite definition grounds event identification in structural, energetic, and kinematic principles, moving beyond the noise-sensitive amplitude thresholds of the classical approach.

**Intensity Classification for Events.**

To characterize the overall strength of an identified MJO event, we classify it based on its mean intensity ($\overline{I_{MJO}}$) over the event's duration: weak ($\overline{I_{MJO}}$<1.5), Medium (1.5≤$\overline{I_{MJO}}$<2.0), Strong (2.0≤$\overline{I_{MJO}}$<2.5), and Extreme ($\overline{I_{MJO}}$≥2.5).

# Data and Code Availability

The atmospheric reanalysis (U850, U200) and OLR data used in this study are publicly available from the following sources:

- NCEP2: https://psl.noaa.gov/data/gridded/data.ncep.reanalysis2.html
- OLR: https://psl.noaa.gov/data/gridded/data.olrcdr.interp.html

The six canonical MJO phase patterns, the code for the real-time MJO index, and the analysis scripts are available on GitHub: https://github.com/zeaccepted/PhysAnchor-MJO-AE

https://doi.org/10.1038/s41467-025-58955-4

Hsu, P., Li, T., & Tsou, C.-H. (2011). Interactions between Boreal Summer Intraseasonal Oscillations and Synoptic-Scale Disturbances over the Western North Pacific. Part I: Energetics Diagnosis*. *Journal of Climate*, *24*(3), 927–941. https://doi.org/10.1175/2010jcli3833.1

Jiang, X., Adames, Á. F., Kim, D.-H., Maloney, E. D., Lin, H., Kim, H.-M., Zhang, C., DeMott, C. A., & Klingaman, N. P. (2020). Fifty Years of Research on the Madden-Julian Oscillation: Recent Progress, Challenges, and Perspectives. *Journal of Geophysical Research: Atmospheres*, *125*(17). https://doi.org/10.1029/2019jd030911

Kanamitsu, M., Ebisuzaki, W., Woollen, J., Yang, S.-K., Hnilo, J. J., Fiorino, M., & Potter, G. L. (2002). NCEP–DOE AMIP-II Reanalysis (R-2). *Bulletin of the American Meteorological Society*, *83*(11), 1631–1644. https://doi.org/10.1175/bams-83-11-1631

Kiladis, G. N., Wheeler, M. C., Haertel, P. T., Straub, K. H., & Roundy, P. E. (2009). Convectively coupled equatorial waves. *Reviews of Geophysics*, *47*(2). https://doi.org/10.1029/2008rg000266

Li, R. C. Y., & Zhou, W. (2013). Modulation of Western North Pacific Tropical Cyclone Activity by the ISO. Part I: Genesis and Intensity. *Journal of Climate*, *26*(9), 2904–2918. https://doi.org/10.1175/jcli-d-12-00210.1

Li, T., Zhao, C., Hsu, P., & Nasuno, T. (2015). MJO Initiation Processes over the Tropical Indian Ocean during DYNAMO/CINDY2011*. *Journal of Climate*, *28*(6), 2121–2135. https://doi.org/10.1175/jcli-d-14-00328.1

Liebmann, B., & Smith, C. A. (1996). Description of a Complete (Interpolated) Outgoing Longwave Radiation Dataset. *Bulletin of the American Meteorological Society*, *77*(6), 1275–1277. https://www.jstor.org/stable/26233278

Madden, R. A., & Julian, P. R. (1971). Detection of a 40–50 Day Oscillation in the Zonal Wind in the Tropical Pacific. *Journal of the Atmospheric Sciences*, *28*(5), 702–708. https://doi.org/10.1175/1520-0469(1971)028%3C0702:doadoi%3E2.0.co;2

Madden, R. A., & Julian, P. R. (1972). Description of Global-Scale Circulation Cells in the Tropics with a 40–50 Day Period. *Journal of the Atmospheric Sciences*, *29*(6),

# Figures

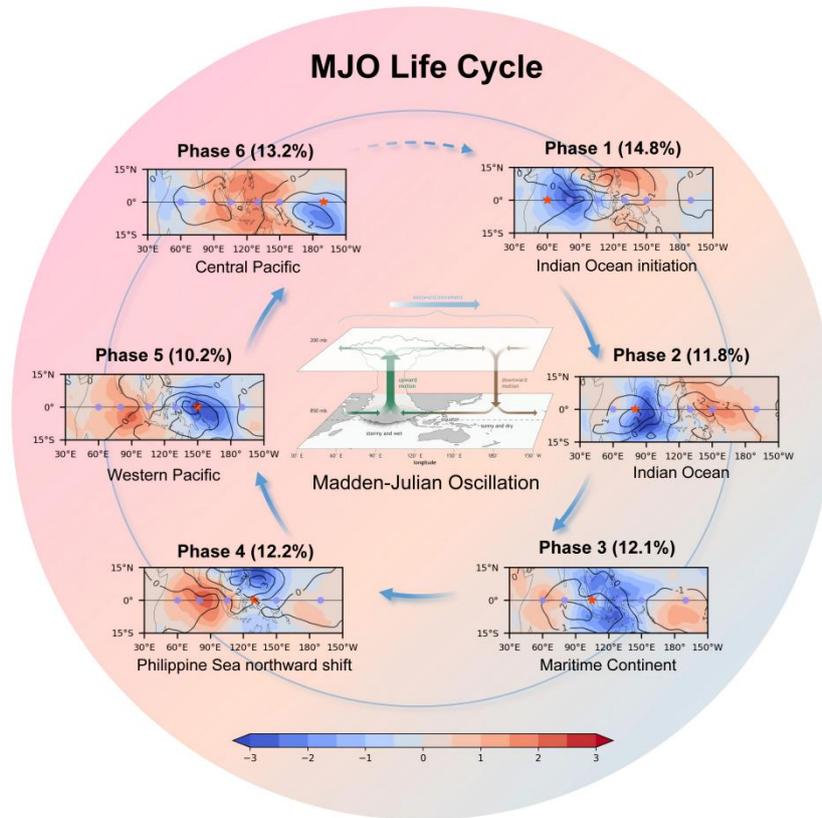

**Figure 1 | The AI-derived anatomical map of the MJO: A six-phase life cycle discovered by clustering in the latent space.**

The six canonical MJO phases were objectively identified by applying a Self-Organizing Map (SOM) to the learned latent vectors, which encode MJO dynamical states such that geometrically close vectors correspond to physically similar and positionally proximate MJO systems. Each phase is represented by its composite spatial structure: shading denotes outgoing longwave radiation (OLR) anomalies, and contours denote 850-hPa zonal wind (U850) anomalies. The red star marks the characteristic system center of the depicted phase; purple dots indicate the centers of the other five phases. The percentage in parentheses is the historical frequency of occurrence for that phase. The phases are ordered by their central longitude to form a complete life cycle. Phases 2 and 4 are newly identified, objective signatures of the long-hypothesized but previously undefined transitional stages: Indian Ocean and Philippine Sea, respectively. The schematic in the center (adapted from NOAA Climate.gov) depicts the canonical eastward-propagating MJO structure for reference.

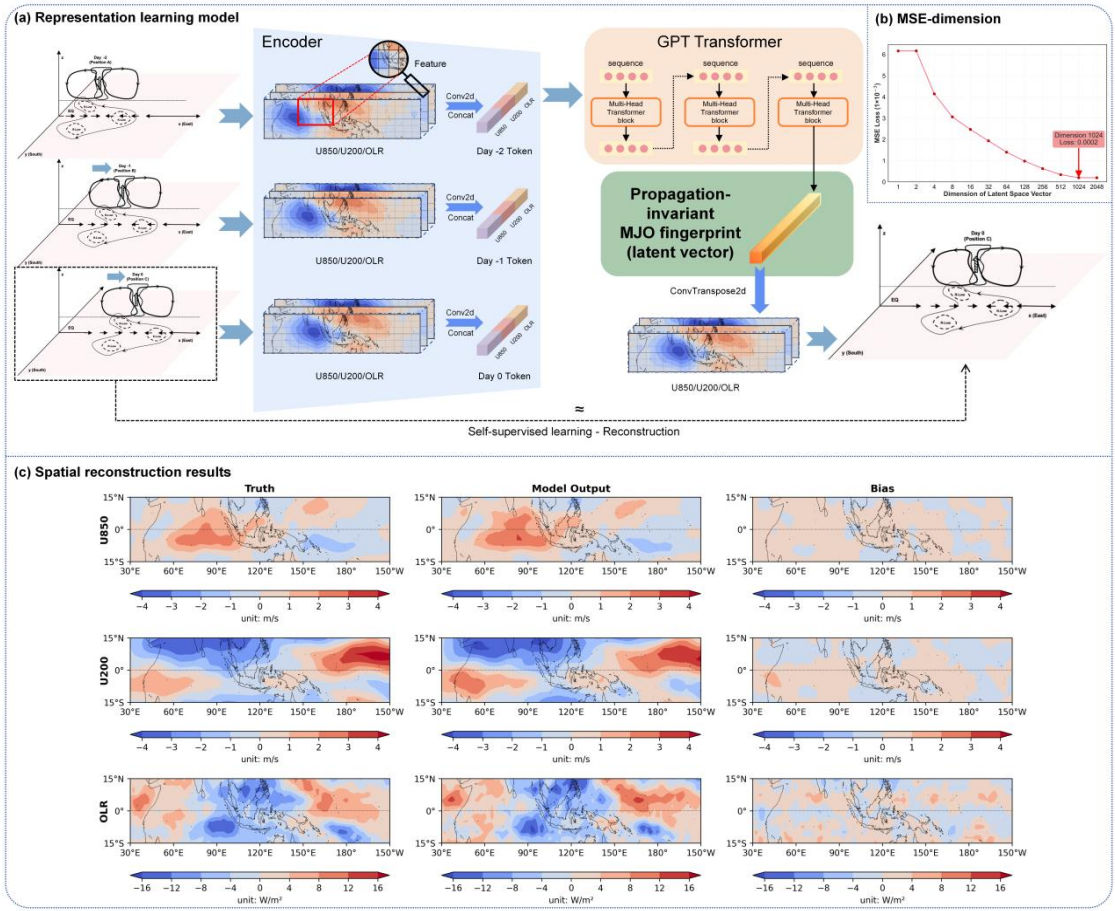

**Fig. 2 | Learning a contextual "MJO fingerprint" using AI.**

(a) Schematic of the Transformer-based encoder-decoder model. The model takes a sequence of three consecutive days of tropical fields—zonal wind at 850 hPa (U850) and 200 hPa (U200), and outgoing longwave radiation (OLR)—and is trained to reconstruct the third day's state. This task forces the encoder to learn to encode the input into a fixed-dimensional latent vector (an "MJO fingerprint") that captures the coherent physical evolution and organizes the latent space such that physically similar and positionally proximate MJO states are mapped close together—thereby overcoming the "propagation penalty." (b) Determination of the optimal fingerprint dimensionality. Reconstruction loss plateaus at 1024 dimensions (red arrow), indicating this latent size suffices to encode the MJO's core dynamical information. (c) Validation of reconstruction fidelity. Spatial composite fields of U850 (top), U200 (middle), and OLR (bottom) from the original data (left column) are nearly indistinguishable from those reconstructed from the 1024-dimensional fingerprints (middle column). The difference fields (right column) are minimal, confirming the fingerprint retains the essential coupled structure of the MJO.

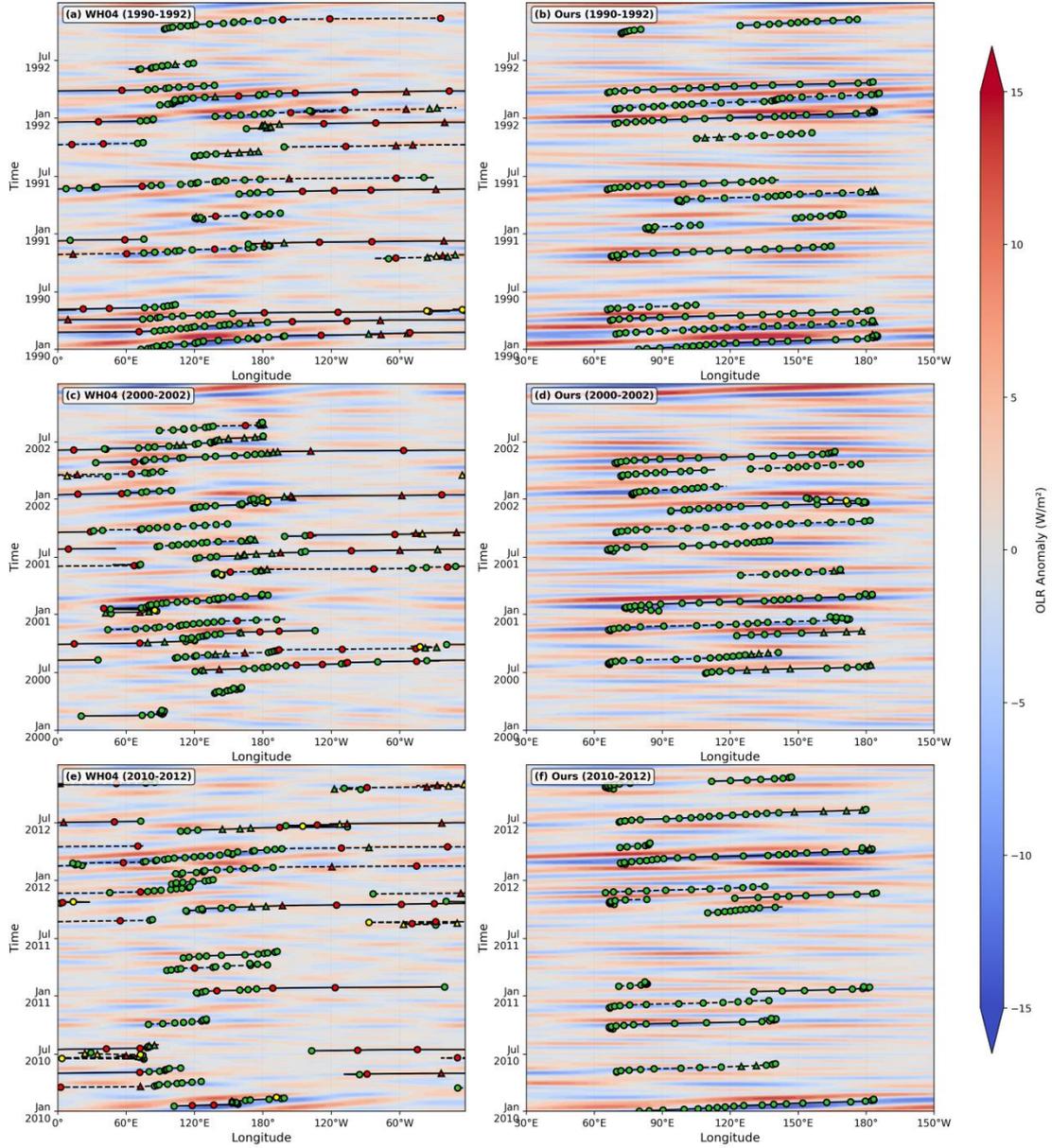

**Figure 3 | Superior physical fidelity and event delineation of the new MJO monitoring framework.**

A multi-decadal comparison of MJO event tracking between the classical WH04 method (left column) and our new framework (right column). All panels are Hovmöller diagrams (longitude-time) of filtered OLR anomalies (shading, 15°S–15°N average). Trajectories show the diagnosed daily longitudinal center of MJO-related convection. Distinct MJO events are separated by solid and dashed lines. Each daily estimate is marked by a symbol whose shape indicates physical consistency with the convective field: circles (center within negative OLR anomalies) and triangles (center within positive OLR anomalies). The fill color of each symbol indicates the diagnosed zonal propagation speed: green (normal eastward, <10° longitude/day), red (fast eastward, ≥10°/day), and yellow (westward). For a spatially continuous comparison, the discrete phases of the WH04 index were mapped to a central longitude. Across all decades, the new method provides physically coherent tracking—closely following convective anomalies, correctly segmenting events, and depicting realistic propagation—while the classical method suffers from frequent misplacement, spurious westward propagation, and erroneous merging of distinct events.

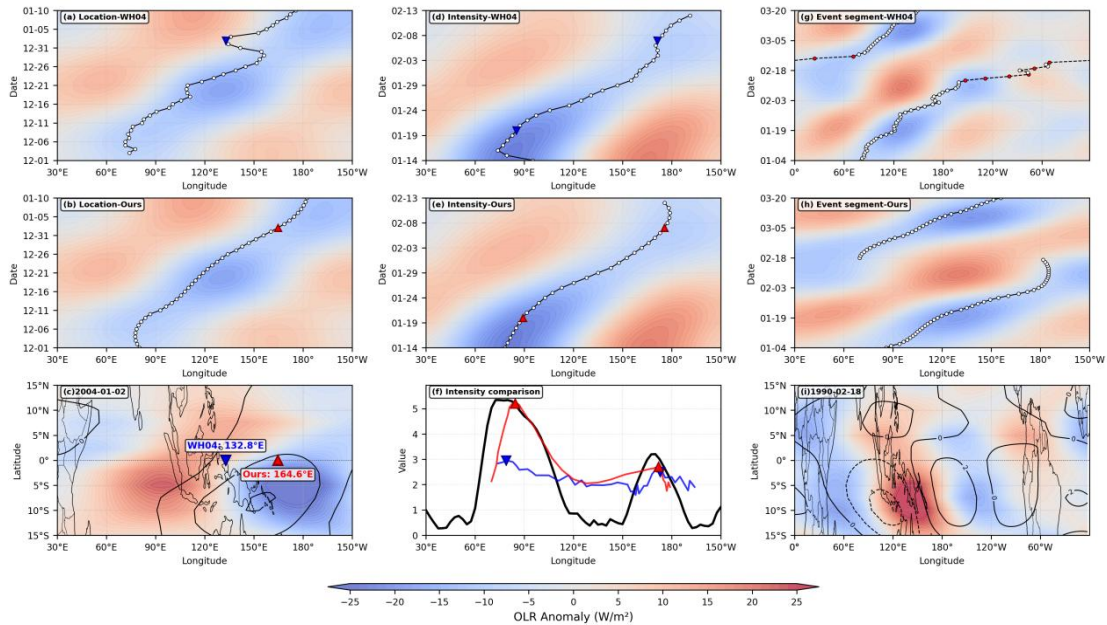

**Figure 4 | Case studies exposing fundamental flaws in classical MJO monitoring and their resolution.**

Three independent MJO events demonstrate specific limitations of the classical WH04 index and the improvements of the new framework. a-c, Problem 1: Spurious location and propagation. For the 2003-2004 event, the Hovmöller diagram (OLR anomalies, shading; U850 anomalies, contours) tracked by the classical index (a) shows a clear, unphysical westward jump (retrogression). The new method (b) correctly tracks consistent eastward propagation. A snapshot (c) on 2004-01-02 shows the classical index (blue triangle) places the MJO center in a convectively suppressed region (positive OLR anomaly), while the new method (red circle) co-locates it with the convective core. d-f, Problem 2: Failure to capture intensity evolution. For the 2005 event, the relative strength of two convective peaks (black line, OLR intensity scaled) with a ratio of ~1.7:1 is correctly reflected by the new intensity index (red line) but is flattened by the classical RMM amplitude (blue line). g-i, Problem 3: Inability to segment concurrent events. During the 1990 event, a convectively suppressed zone separates two distinct MJO systems. The classical index (g) merges them into one prolonged event. The new method (h) applies a physically-based disambiguation rule, identifying the dominant event until the new, initiating system in the Indian Ocean surpasses it, leading to a clean segmentation into two life cycles. The snapshot on the transition date (i) confirms the coexistence of two systems; the new method attributes the location to the stronger, pre-existing system in the central Pacific.

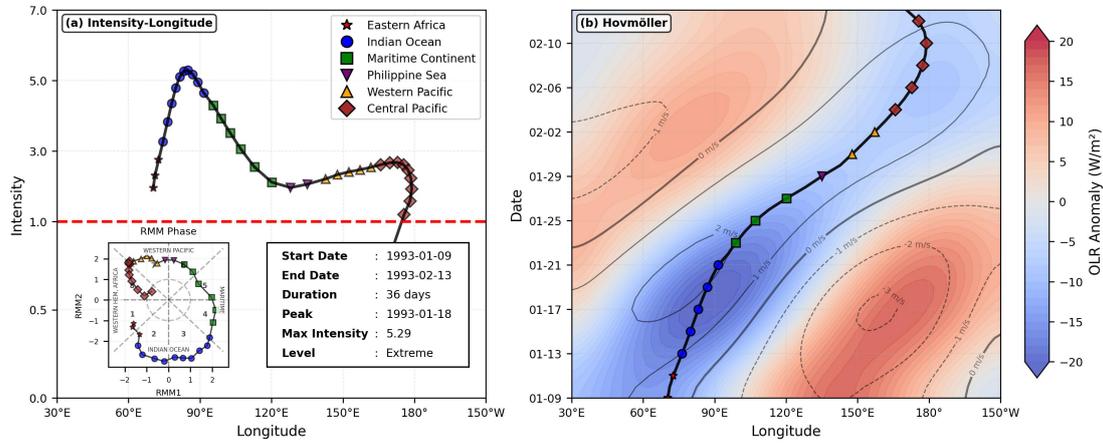

**Figure 5 | A standardized, physics-based template for comprehensive MJO lifecycle analysis.**

The proposed template synthesizes the output of the new framework into a unified visualization for diagnosing individual MJO events. a, Phase and intensity evolution. The primary panel plots the daily MJO state in a longitude-intensity space, with symbols and colors corresponding to the six canonical phases (Fig. 1), tracing the event's coherent progression. The inset shows the conventional WH04 RMM phase-space diagram for comparison. Key event metadata are summarized at right. b, Dynamical and convective context. A Hovmöller diagram displays filtered OLR anomalies (shading) and 850-hPa zonal wind (U850) anomalies (contours) averaged over 15°S–15°N. The daily MJO location and its dominant phase (using the same symbols as in panel a) are overlaid, demonstrating precise tracking of the convective core. This template integrates information on location, intensity, structural phase, and environmental context, providing a more complete and physically consistent representation than the classical phase diagram and serving as a proposed standard for future MJO analysis.

# Supplementary

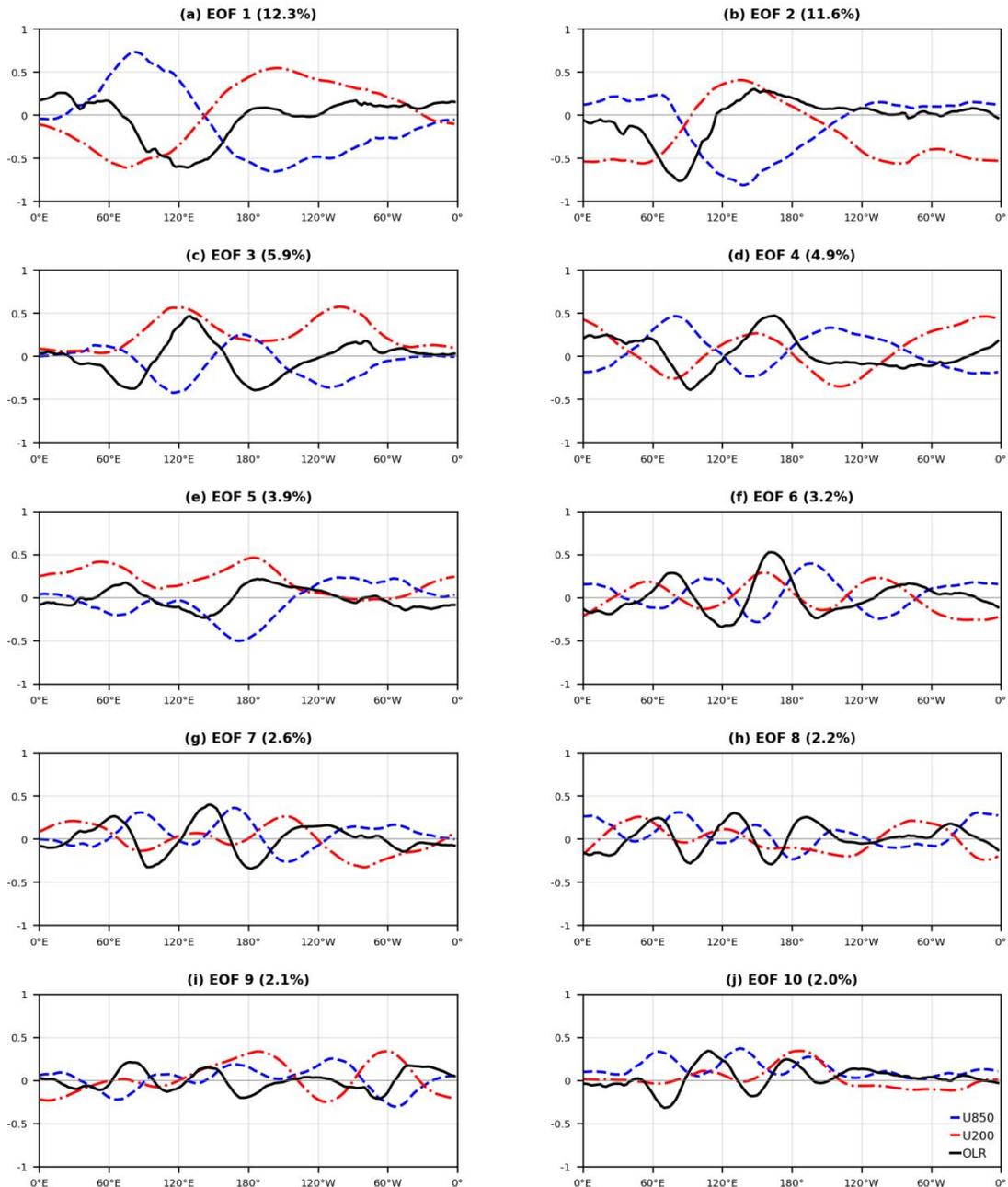

**Figure S1 | Inadequacy of expanding the classical linear basis for representing the MJO.**

The first ten multivariate Empirical Orthogonal Function (EOF) patterns derived from the classical WH04 methodology, applied to 20–90-day filtered, near-equatorially averaged (15°S–15°N) fields of U850, U200, and OLR. The percentage of explained variance for each mode is shown in parentheses. While the first two modes (a, b) form the basis of the WH04 RMM index, modes 3–10 (c–j) collectively explain only ~26% additional variance. Critically, these higher-order modes exhibit multi-celled zonal structures that deviate from the fundamental, planetary-scale zonal wavenumber-1 signature of the MJO, demonstrating that simply retaining more EOF modes cannot resolve the inherent limitations of the linear framework for capturing the MJO's complex life cycle.

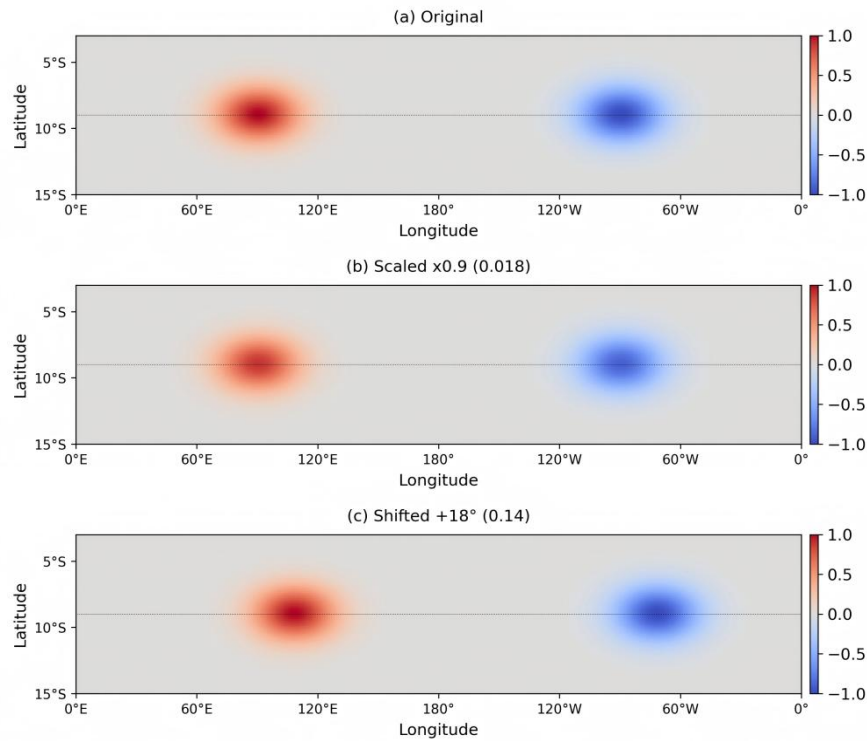

**Figure S2 | The "propagation penalty": Why clustering in raw data space fails for the MJO.**

An idealized experiment demonstrating the inherent bias of geometric similarity metrics (e.g., Euclidean distance) against a propagating phenomenon. A simplified MJO-like convective-circulation dipole is used as a benchmark state (a). The distance is calculated to the same dipole after a simulated 3-day intensity weakening (amplitude scaled to 90%, b) and after a simulated eastward propagation (zonal shift of ~18° longitude, c). Euclidean distance is vastly more sensitive to positional shift (0.14) than to the structural intensity change (0.018). This fundamental bias means any clustering algorithm applied directly to raw atmospheric fields would primarily separate states based on their zonal location, treating the same MJO structure at different longitudes as distinct classes. This "propagation penalty" explains why traditional approaches cannot objectively identify the recurrent, propagating entity that is the MJO, necessitating the learning of a representation that overcomes this bias by mapping physically similar and positionally proximate states to nearby points in a latent space.

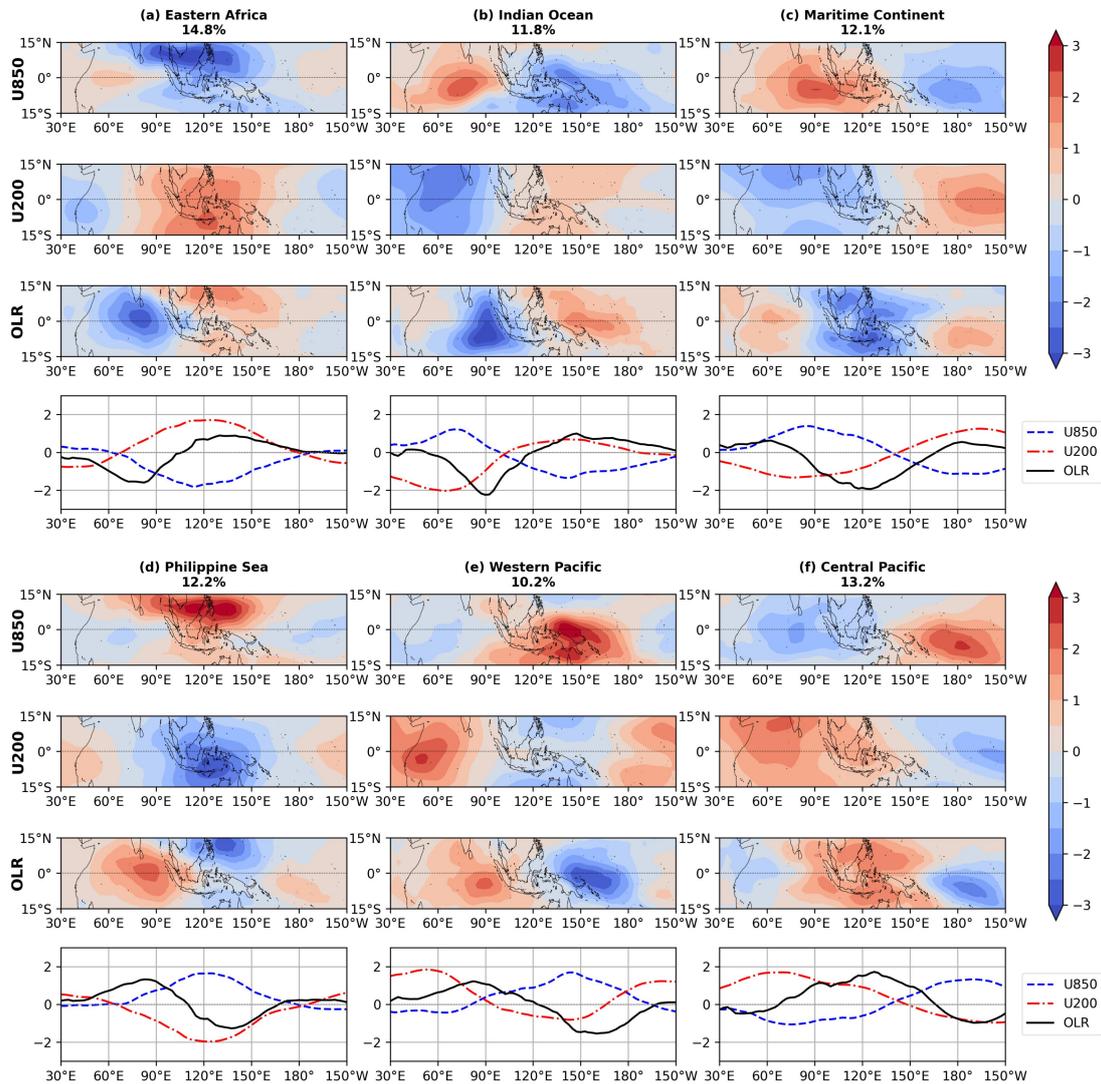

**Figure S3 | Detailed spatial structure and occurrence frequency of the six canonical MJO phases.**

The six canonical MJO phases objectively identified by Self-Organizing Map (SOM) clustering of the latent vectors, ordered from west to east by the longitude of their convective center. For each phase, the composite spatial anomaly patterns of U850, U200, and OLR are shown (top three panels). The bottom panel for each phase displays the corresponding meridionally averaged (15°S–15°N) anomalies, enabling direct comparison with classical EOF patterns. The percentage in each subplot title indicates the historical frequency of occurrence for that phase. This detailed view complements Figure 1, confirming that each phase represents a coherent, recurrent dynamical state within the MJO's life cycle.

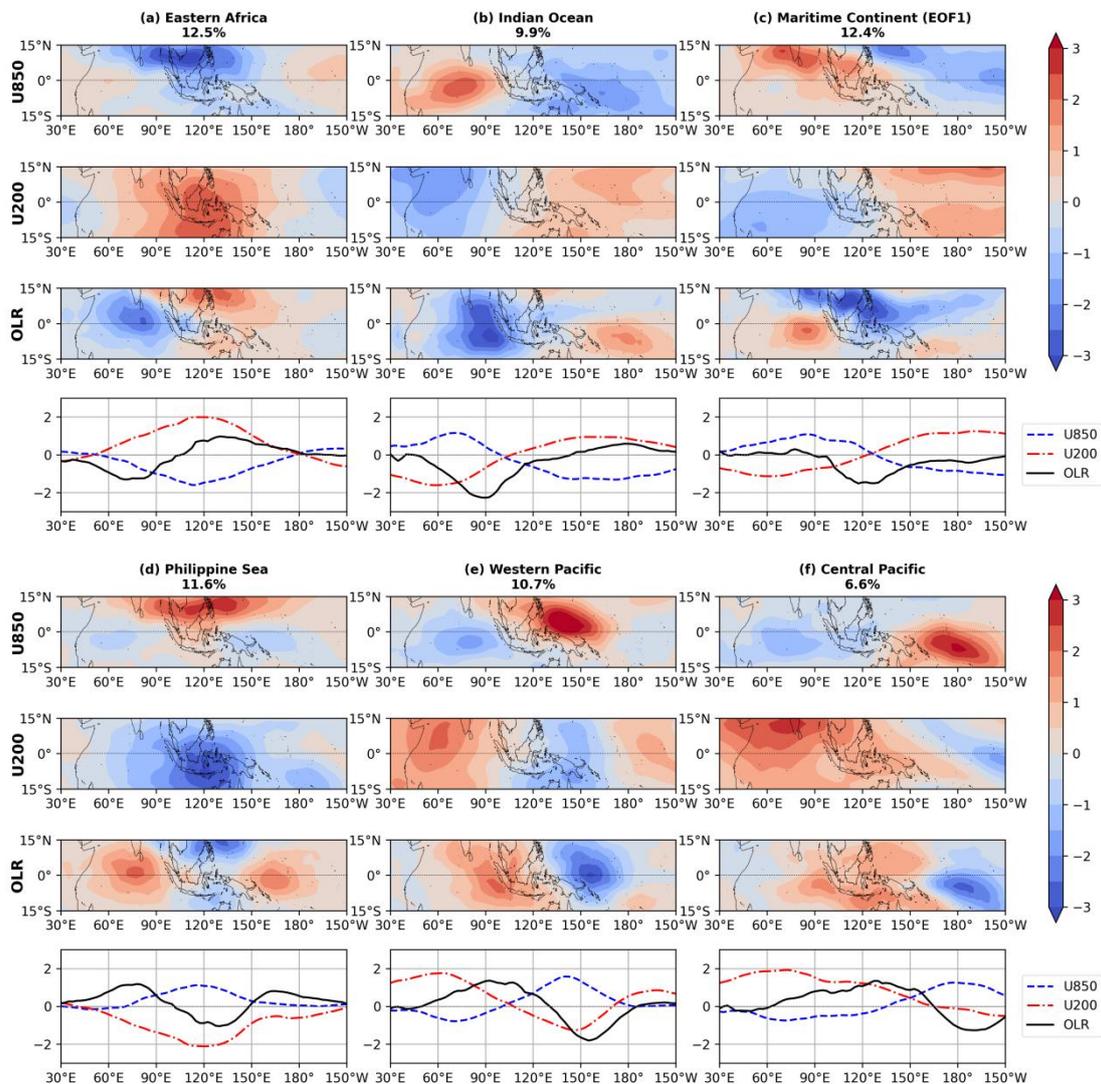

**Figure S4 | Robustness validation: Consistent MJO phases emerge from linear K‑means clustering.**

The six primary MJO phases identified by applying linear K‑means clustering to the same learned latent vectors (which encode MJO states such that geometrically close vectors correspond to physically similar and positionally proximate systems). The layout is identical to Figure S3, showing the spatial anomaly patterns of U850, U200, and OLR for each phase, followed by their meridional averages. The high degree of consistency in spatial structures and the sequential progression of convective centers with the phases derived from the nonlinear SOM algorithm (Fig. S3) demonstrates that the six-phase life cycle is a robust and objective property of the MJO dynamics captured by our latent representation, not an artifact of the specific clustering methodology.

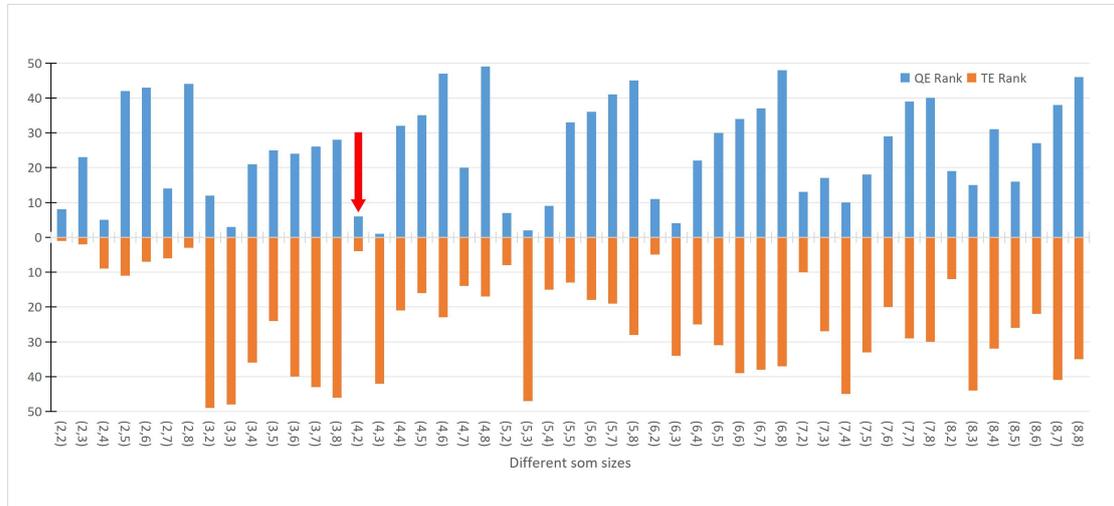

**Figure S5 | Objective determination of the optimal cluster number for the Self-Organizing Map (SOM).**

Quantization Error (QE, blue bars) and Topographic Error (TE, red bars) for SOM configurations ranging from 2x2 to 8x8 nodes, applied to the latent vectors. The combined error is minimized for the (4,2) topology (8 nodes, red arrow), which was therefore selected to define the canonical MJO phases. This data-driven choice provides an objective basis for the subsequent clustering analysis that revealed the six-phase life cycle.

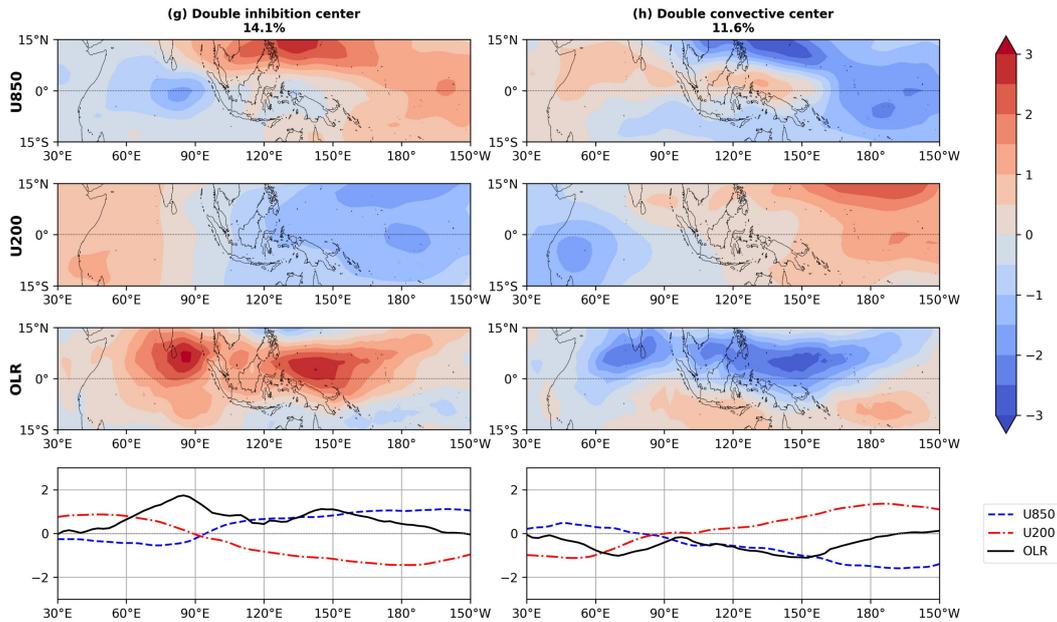

**Figure S6 | Two non-canonical clusters excluded from the MJO phase taxonomy.**

Spatial structures of the two clusters identified by the Self-Organizing Map (SOM) that were excluded from the set of six canonical MJO phases. They are displayed in the same format as the canonical phases in Figure S3, showing anomaly patterns for U850, U200, and OLR, along with their meridional averages. These patterns lack the definitive planetary-scale, wavenumber-1 structure and the coupled convective-circulation signature that physically define the MJO. Their exclusion underscores the objective, physically grounded criteria used in the selection process, ensuring the final taxonomy captures only the core dynamical regimes of the phenomenon.

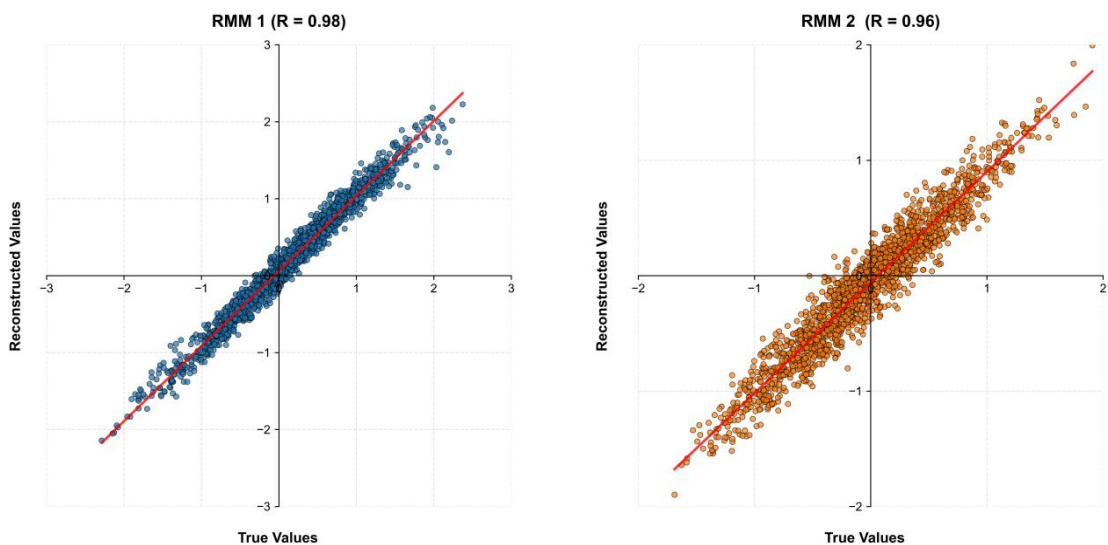

**Figure S7 | Validation of the latent representation from an MJO monitoring perspective.**

Scatter plots comparing the classical RMM indices calculated from the original observational data (x-axis) vs. those calculated from the model-reconstructed data (y-axis). Left, RMM1; Right, RMM2. The high correlation coefficients (R > 0.95) indicate that the reconstruction process, and thus the learned latent representation, preserves virtually all the phase information that defines the MJO within the classical monitoring framework. This confirms that the essential MJO signal has been successfully encoded into a latent space whose geometry reflects dynamical similarity, complementing the spatial validation shown in Fig. 2c.

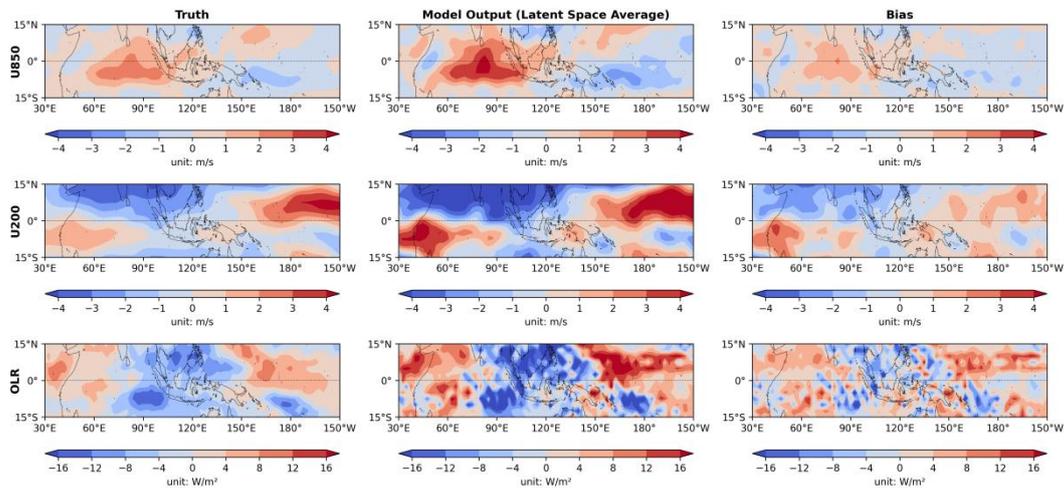

**Figure S8 | Validation of the latent space as a composable vector space for clustering.**

Spatial composites for each input variable (U850, top row; U200, middle row; OLR, bottom row). The left column shows composites derived by averaging the original observational fields. The middle column shows fields reconstructed from the average of all corresponding daily latent vectors. The right column shows the difference. The reconstructed composites successfully recover the spatial patterns of the originals, confirming that averaging in the latent space yields a physically plausible result. The slight overestimation in magnitude is attributable to the nonlinear decoder. This demonstrates that the latent vectors reside in a space where linear operations (like computing a centroid for clustering) are meaningful, providing the mathematical foundation for the subsequent identification of canonical MJO phases.

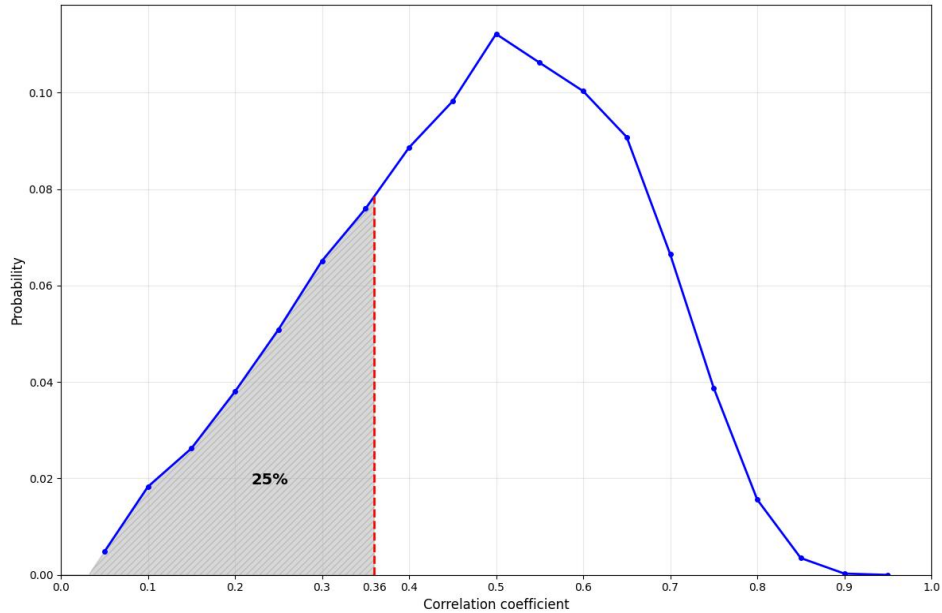

**Figure S9 | Objective determination of the pattern-similarity threshold for MJO event identification.**

Probability density function of the daily maximum spatial pattern correlation between the observed atmospheric fields (U850, U200, OLR) and the six canonical MJO phases for the period 1981–2020. The 25th percentile of this historical distribution (correlation = 0.36, vertical dashed line) is selected as the classification threshold (red shaded area). This statistically objective cutoff is chosen to maintain physical consistency: it systematically filters out a proportion of days comparable to the ~25% of non-canonical states discarded during the Self-Organizing Map clustering, thereby providing a unified criterion for distinguishing days with a structured MJO signal from those dominated by unstructured variability in the monitoring algorithm.